\documentclass[fleqn,usenatbib]{mnras}

\usepackage[T1]{fontenc}
\usepackage{ae,aecompl}

\usepackage{graphicx}	
\usepackage{amsmath}	
\usepackage{amssymb}	
\usepackage{newtxtext,newtxmath}

\title[Perseus Cluster UDGs]{Ultra-Diffuse Galaxies in the Perseus Cluster: Comparing Galaxy Properties with Globular Cluster System Richness}

\author[J. S. Gannon et al.]{Jonah S. Gannon$^{1}$\thanks{E-mail: jgannon@swin.edu.au}, Duncan A. Forbes$^{1}$, Aaron J. Romanowsky$^{2,3}$,  Anna Ferr\'e-Mateu$^{4,1}$, 
\newauthor{Warrick J. Couch$^{1}$, Jean P. Brodie$^{1,3}$, Song Huang$^{5}$, Steven R. Janssens$^{6}$}
\newauthor{and Nobuhiro Okabe$^{7}$}
\\
$^{1}$ Centre for Astrophysics and Supercomputing, Swinburne University, John Street, Hawthorn VIC 3122, Australia
\\
$^{2}$ Department of Physics and Astronomy, San Jos\'e State University, One Washington Square, San Jose, CA 95192, USA
\\
$^{3}$ University of California Observatories, 1156 High Street, Santa Cruz, CA 95064, USA
\\
$^{4}$ Institut de Ci\'encies del Cosmos (ICCUB), Universitat de Barcelona (IEEC-UB), Barcelona 08028, Spain
\\
$^{5}$ Department of Astrophysical Sciences, Princeton University, Princeton, NJ 08544, USA
\\
$^{6}$ Department of Astronomy \& Astrophysics, University of Toronto, 50 St. George Street, Toronto, ON M5S 3H4, Canada
\\
$^{7}$ Department of Physics, Hiroshima University, 1-3-1 Kagamiyama, Higashi-Hiroshima, Hiroshima 739-8526, Japan
}

\date{Accepted XXX. Received YYY; in original form ZZZ}

\pubyear{2021}

\begin{document}
\label{firstpage}
\pagerange{\pageref{firstpage}--\pageref{lastpage}}
\maketitle

\begin{abstract}
It is clear that within the class of ultra-diffuse galaxies (UDGs) there is an extreme range in the richness of their associated globular cluster (GC) systems. Here, we report the structural properties of five UDGs in the Perseus cluster based on deep Subaru / Hyper Suprime-Cam imaging. Three appear GC-poor and two appear GC-rich.  One of our sample, PUDG\_R24, appears to be undergoing quenching and is expected to fade into the UDG regime within the next $\sim0.5$ Gyr. We target this sample with Keck Cosmic Web Imager (KCWI) spectroscopy to investigate differences in their dark matter halos, as expected from their differing GC content. Our spectroscopy measures both recessional velocities, confirming Perseus cluster membership, and stellar velocity dispersions, to measure dynamical masses within their half-light radius. We supplement our data with that from the literature to examine trends in galaxy parameters with GC system richness. We do not find the correlation between GC numbers and UDG phase space positioning expected if GC-rich UDGs environmentally quench at high redshift. We do find GC-rich UDGs to have higher velocity dispersions than GC-poor UDGs on average, resulting in greater dynamical mass within the half-light radius. This agrees with the first order expectation that GC-rich UDGs have higher halo masses than GC-poor UDGs. 
\end{abstract}




\section{Introduction}

\begin{figure*}
    \centering
    \includegraphics[width = 0.99 \textwidth]{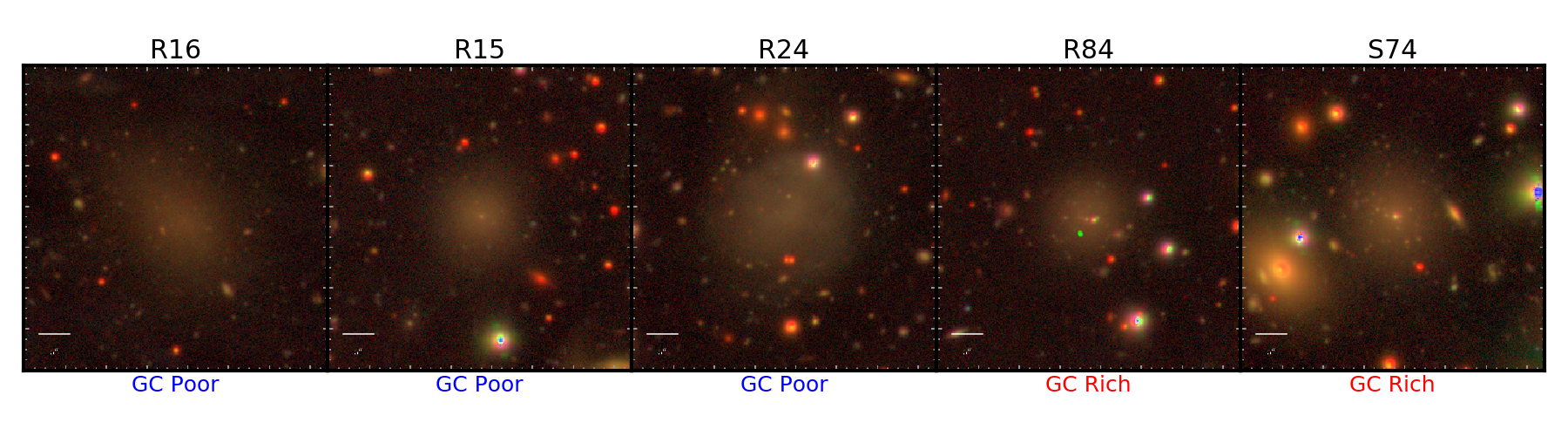}
    \caption{Composite colour cutouts around each UDG from the Subaru / Hyper-Suprime Cam $g$,$r$,$i$-band data. North is up and east is left. The white bar in each image indicates a 5'' scale. Three UDGs, R15, S74 and R84, have central compact sources that are likely galaxy stellar nuclei. We arrange the UDGs, left to right, in order of increasing GC richness. The UDGs S74 and R84 appear to contain additional compact sources and are classified as GC-rich UDGs. Conversely, R15, R16 and R24 appear associated with fewer compact sources and are classified as GC-poor.}
    \label{fig:gc_rich_poor}
\end{figure*}

While the existence of large, low surface brightness galaxies has been known for many decades (e.g., \citealp{Disney1976, Sandage1984, Bothun1987, Impey1988, Impey1997}), the work of \citet{vanDokkum2015} has reignited interest in their study. The latter study identified 47 low surface brightness within the Coma cluster and established the working definition of `ultra-diffuse galaxy' (UDG) based on two criteria: half-light radius, $R_{\rm e} > 1.5$ kpc, and surface brightness, $\mu_{0,g}>24\ \mathrm{mag\ arcsec^{-2}}$ \citep{vanDokkum2015}.

Many UDGs have been shown to host extensive globular cluster (GC) systems (e.g., \citealp{Beasley2016, Beasley2016b, vanDokkum2017, vanDokkum2018b, Amorisco2018, Lim2018, Toloba2018, Prole2019, Forbes2019, Roman2019, lim2020, Somalwar2020, Montes2020, Montes2021, Shen2021, Muller2021}). It has been demonstrated that dense environments can impact a galaxy's GC system, increasing the GC system richness relative to the stellar mass (i.e., the GC specific frequency; \citealp{Peng2008, Liu2016, Lim2018}). Early indications are that this is also true for UDGs, particularly in the Coma cluster \citep{Lim2018, Forbes2020, Somalwar2020}. Three possible explanations for this effect have been proposed in the literature: 1) increased GC production due to the earlier and faster formation times of cluster galaxies, 2) decreased GC destruction due to earlier quenching times, and 3) decreased field star formation due to earlier quenching times (see e.g., \citealp{Peng2008, Carlsten2021}). 

Interestingly, these explanations align with some proposals for forming the large sizes and/or low surface brightnesses of UDGs. Rapid star formation timescales will introduce a sharp impulse of energy into the system, which has been studied as a possible avenue for creating UDG's large sizes (e.g., \citealp{DiCintio2017, Chan2018, Martin2019, Jiang2019}). Early infall times will quench star formation resulting in a passively evolving stellar population that will slowly decrease in surface brightness (e.g, \citealp{Yozin2015, Roman2017b, Chan2018, Jiang2019, Sales2020, Tremmel2020}). Additionally, UDGs that fall in earlier will spend more time being subject to the cluster's tidal forces, which can play an important role in heating and expanding UDG stellar populations, giving rise to their large sizes \citep{Yozin2015, Jiang2019, Martin2019, Sales2020, Carleton2018, Carleton2021}. Other mechanisms (e.g., high halo spin; \citealp{Amorisco2016, Rong2017, Pina2020}) and combinations of mechanisms (e.g., stellar feedback and quenching, \citealp{Chan2018} or stellar feedback and tides, \citealp{Jiang2019}) have also been considered.

Up until now, simulations of UDG formation have primarily focused on the formation of their stellar body (see e.g., \citealp{Yozin2015, DiCintio2017,  Chan2018, Carleton2018, Liao2019, Jiang2019, Martin2019, Sales2020, Tremmel2020, Wright2021}) but very few have focused on modelling their associated GC systems. Large cosmological simulations of galaxy formation do not (yet) have the resolution to properly resolve GC formation. Probing galaxy GC formation requires either the implementation of specialised simulations for single galaxies (e.g., \citealp{Chowdhury2020}) or additional sub-grid/semi-analytic modelling within the simulation (e.g., \citealp{Carleton2021, Doppel2021}). To date, only one work has applied such semi-analytic models to a cosmological simulation to investigate GC formation in UDGs. \citet{Carleton2021} used a semi-analytic model for GC formation along with the IllustrisTNG simulations to find that UDGs with large GC populations likely formed through a combination of fast, high-redshift star formation and tidal heating in a cluster environment.

The effects of external tidal fields acting on a UDG is highly dependent on the structure of its dark matter halo (e.g., cusp \textit{vs.} core and total mass). GCs are a known indicator of total dark matter halo mass for normal galaxies \citep{Spitler2009, Harris2013, Harris2017, Burkert2020}. Currently this relation has only one independent confirmation for UDGs coming from a resolved mass profile for the UDG Dragonfly 44 \citep{vanDokkum2019b, Forbes2021}. Using the \citet{Burkert2020} relation, \citet{Forbes2020} studied the GC systems of UDGs in the Coma cluster and suggested the existence of two types of UDG dark matter halo. One type is GC-poor with dwarf-like dark matter halos ($M_{\mathrm{Halo}} \approx 10^{10} \mathrm{M_{\odot}}$) and the other is GC-rich with massive, $10^{11} - 10^{12} \mathrm{M_{\odot}}$ dark matter halos. Many authors propose these GC-rich UDGs may be examples of ``failed galaxies" that were rapidly quenched after a period of early, rapid star formation \citep{vanDokkum2017, Forbes2020, Villaume2021}. The population of GC-poor UDGs is expected to be more mixed, resembling ``puffy dwarf'' galaxies - normal dwarfs that have been ``puffed up" through a variety of processes to larger sizes and low surface brightnesses. We note the existence of tidal UDGs (e.g., \citealp{Ogiya2018, Collins2020, Roman2021, Iodice2021}) and other low dark matter UDGs which will not fit into these two types but this is beyond the scope of this work.

Mass measurements coming from stellar, and GC velocity dispersions, for many GC-rich UDGs confirm their dark matter dominated nature \citep{vanDokkum2016, Beasley2016, vanDokkum2017, Toloba2018, vanDokkum2019b, Martin-Navarro2019, Forbes2021, Gannon2020, Gannon2021}. When used in comparison to a total halo mass estimate coming from GC numbers they can be used to infer basic halo properties (e.g., cusp \textit{vs.} core) \citep{Toloba2018, Gannon2020, Gannon2021, Forbes2021}.


Also of importance in studying external environmental effects on UDGs, such as early quenching, are their position -- velocity locations within their environment \citep{Alabi2018}. To first order, galaxies populate distinct regions in phase space which are dependent on their infall time \citep{Rhee2017}. UDGs that have spent longer in the higher density environment will have quenched earlier and will have been subjected to its tidal field for a longer period (see section 5.2 of \citealp{Martin2019} for a discussion).

\begin{figure*}
    \centering
    \includegraphics[width = 0.95 \textwidth]{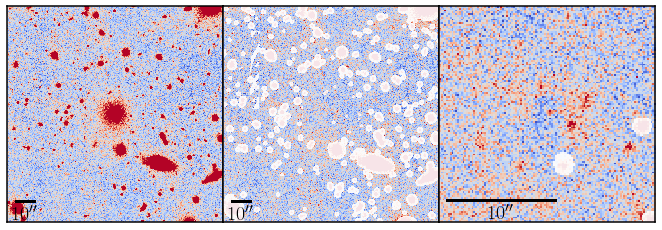}
    \caption{An example of our UDG \texttt{imfit} fitting for R15. \textit{Left panel:} Our original Hyper Suprime-Cam data. \textit{Middle panel:} The residuals of our fit with areas masked in the fitting overplotted (white regions). \textit{Right panel:} A zoom-in at the location of the UDG in the residual image. We plot with `asinh' image stretching in this panel to make the residual structure more evident. For each panel North is up and East is left. A 10'' bar is also included in each panel.}
    \label{fig:r15_imblock}
\end{figure*}

In this work, we present a differential analysis of five UDGs in the Perseus cluster that were visually identified as having different degrees of GC richness. The Perseus cluster (mass = $1.2 \times10^{15}\ \mathrm{M_{\odot}}$, mean $V_{r} = 5258\ \mathrm{km\ s^{-1}}$, $\sigma = 1040\ \mathrm{km\ s^{-1}}$ and $R_{200} = 2.2$ Mpc; \citealp{Aguerri2020}) represents a good analogue of the Coma cluster (mass = $2.7\times10^{15}\ \mathrm{M_{\odot}}$, mean $V_{r} = 6943\ \mathrm{km\ s^{-1}}$, $\sigma = 1031\ \mathrm{km\ s^{-1}}$ and $R_{200} = 2.9$ Mpc; \citealp{Alabi2018}) where UDGs have already been studied extensively. It is, however, closer ($D_{\rm Perseus} \approx 75$ Mpc; $D_{\rm Coma} \approx 100$ Mpc), allowing better study of its UDG population. We therefore use integral field spectroscopy, in line with previous UDG studies (e.g., \citealp{Emsellem2018, vanDokkum2019b, Danieli2019, Martin-Navarro2019, Muller2020, Gannon2020, Gannon2021, Forbes2021}), to look for differences in their dark matter mass as expected by their GC numbers. We use their recessional velocities to place them in the phase space diagram for the Perseus cluster in order to explore differences in their infall times. We compare our dynamical mass measures to the halo mass expected from their GC systems in order to explore the underlying properties of their dark matter halo (e.g., cusp \textit{vs.} core and concentration). Finally, we supplement our Perseus UDG observations with those from the literature in order to explore correlations of galaxy properties with GC system richness.

The structure of the paper is as follows: In Section \ref{sec:HSC_Data} we present our Perseus UDG sample and its photometric properties based on Hyper Suprime-Cam imaging. In Section \ref{sec:KCWI_Data} we present and analyse newly acquired Keck Cosmic Web Imager (KCWI) spectroscopy of our Perseus UDG sample. We measure their recessional velocities and stellar velocity dispersions. We discuss our data in the context of UDG formation in Section \ref{sec:Discussion}. Here, we place particular emphasis on the difference between the GC-rich and GC-poor UDGs in our sample. In Section \ref{sec:Conclusions} we present the concluding remarks of our study.

\section{Hyper Suprime-Cam Imaging} \label{sec:HSC_Data}

\begin{table*}
\begin{tabular}{ccccccccccc}
\\ \hline
Name & RA & Dec & $R_{\rm clust}$ & $R_{\rm e, \mathrm{maj}}$ & $ \langle\mu\rangle_{e, g}$ & $b$/$a$ & $M_{g}$ & $M_{\star}$ & $g-i$ & GC Richness \\
 PUDG\_ & [Deg] & [Deg] & [Mpc] & [''/kpc] & [$\mathrm{mag\ arcsec^{-2}}$] &  & [mag] & [$\mathrm{\times 10^{8}\ M_{\odot}}$] & [mag] & \\ \hline
R16       & 49.65202 & 41.19229 & 0.51 & 11.6/4.2 (0.06) & 25.70 (0.01) & 0.70 & -15.60 (0.01) & 5.75 & 1.04 & Poor \{1\} \\
R15$^{*}$ & 49.26584 & 41.24856 & 0.76 & 6.8/2.5=  (0.02) & 25.13 (0.02) & 0.97 & -15.35 (0.02) & 2.59 & 0.85 & Poor \{1\} \\
R24       & 49.64830 & 41.80890 & 0.49 & 9.8/3.6  (0.08) & 24.75 (0.02) & 0.81 & -16.34 (0.01) & 3.91 & 0.68 & Poor \{5\}\\
R84$^{*}$ & 49.35375 & 41.73929 & 0.66 & 5.6/2.0  (0.01) & 24.98 (0.01) & 0.97 & -15.10 (0.01) & 2.20 & 0.87 & Rich \{28\}\\
S74$^{*}$ & 49.29835 & 41.16787 & 0.78 & 10.5/3.8 (0.03) & 25.12 (0.02) & 0.86 & -16.19 (0.01) & 7.85 & 0.96 & Rich \{30\}\\
\hline
\end{tabular}
\caption{A summary of the imaging properties of our Perseus UDG sample. Columns from left to right are as follows: 1) Designation used throughout this work; 2) Right Ascension; 3) Declination; 4) Distance from cluster centre; 5) Semi-major half-light radius ($g$-band); 6) Average surface brightness within half-light radius (extinction corrected); 7) Axial ratio ($g$-band); 8) Absolute $g$-band magnitude (extinction corrected); 9) Stellar mass; 10) $g-i$ band colour (extinction corrected) and 11) GC system richness. For GC system richness we include the number of GCs used in this work in \{curly brackets\} after the rich/poor descriptor. For S74 and R24 these are assumed values while the remaining UDGs have preliminary GC richness measurements from the upcoming work of Janssens et al. (in prep.). We note these GC richness measures may be subject to slight revision when published by Janssens et al. (in prep.), however they are unlikely to change from GC-rich to GC-poor (and vice versa). The measurement of photometric properties from HSC imaging is described in Section \ref{sec:HSC_Data}. UDGs that have nuclei and were fitted with a Gaussian + S\'ersic profile are indicated with a `$^*$'. The remaining UDGs were fitted with a single S\'ersic profile. A distance of 75 Mpc for the Perseus cluster is assumed. Absolute magnitudes and surface brightnesses are corrected for $g$-band Galactic extinction (min 0.495 mag; max 0.685 mag) using \citet{Schlafly2011}. When relevant, uncertainties are given in (brackets) after values. }
 \label{tab:imaging}
\end{table*}

\subsection{HSC Acquisition and Sample Selection} \label{sec:vis_identify}
The Subaru Hyper Suprime-Cam (HSC; \citealp{HSC}) data used in this work were acquired on the night 2014, September 24. This program targeted the Perseus cluster with an image covering a 1.5-degree diameter field of view in three filter bands ($g$,$r$,$i$-bands). These data were reduced via the standard HSC pipeline \citep[ver 4.0.5;][]{HSC_pipe}. While these data were acquired in three filter bands we use the $g$-band primarily in this work. The final $g$-band data stack has a total exposure time of 2160s. Using the definition of \citet{Roman2020} it reaches a surface brightness depth ($3\sigma$, 10$\times$10'' box) of 28.3 $\mathrm{mag\ arcsec^{-2}}$ with 0.8'' seeing.

Initial target selection was drawn from archival CFHT/MegaCam $g$-band imaging covering a 1$\times$1 degree region centred on the Perseus cluster.  A visual comparison was made with the \citet{Wittman2017} catalogue of galaxies to identify new candidate large low surface brightness galaxies. These were given the designations R1 through R148. The more interesting candidates were initially analysed using \texttt{GALFIT} \citep{Peng2010} to derive basic photometric properties. Subsequently, an expanded catalogue of Perseus LSB galaxies was assembled based on the above HSC $g$,$r$,$i$-band imaging, and with an `S'-prefix numbering system.

It was our intent to create a sample of UDGs broadly controlled for size, surface brightness and distance from the cluster centre in order to examine differences primarily based on their GC content. We therefore selected 2 GC-poor (PUDG\_R15 and PUDG\_R16) and 2 GC-rich UDGs (PUDG\_S74 and PUDG\_R84) from the initial \texttt{GALFIT} fitting of these catalogues with closely matched sizes, surface brightnesses and cluster--centric radii (see Table \ref{tab:imaging}). We determined GC richness visually through inspection of compact sources around the galaxies. We compared compact source detections within the UDG half-light radius to nearby, equivalently sized offset regions looking for compact source overdensities on the UDG. We perform this nearby assessment to accurately establish a background level in order to remove potential contamination from GCs associated with nearby bright galaxies and intra-cluster GCs.

We have confirmation of our qualitative assessment of GC richness from \textit{Hubble Space Telescope} (HST) imaging for three of these UDGs: PUDG\_R15, PUDG\_R16 and PUDG\_R84 \citep{Harris2020}. Briefly, \citet{Harris2020} used two band HST data to create a GC catalogue with a colour and point source based selection of GC candidates. These data were observed as part of the same HST program and have approximately the same depth. Offset sky regions were used to establish background contaminant levels and artificial stars were used to  estimate completeness. For full details of the imaging and GC selection see \citet{Harris2020}. Full GC counts for Perseus UDGs from the HST data, including the three UDGs which overlap with this work, will be provided in Janssens et al. (in prep.).

We refer to galaxies as `GC-rich' when they host a GC system richer than 20 GCs. This corresponds to a halo mass of $10^{11}\ \mathrm{M_\odot}$ \citep{Burkert2020} which is more massive than what is expected for the majority of UDGs forming as `puffy dwarfs' and likely indicates a ``failed galaxy'' UDG. UDGs estimated to host a GC system of less than 20 GCs are referred to as `GC-poor'.

The sample of four Perseus cluster UDGs was then supplemented with a fifth UDG, PUDG\_R24 that is located in the cluster outskirts. It is blue in colour ($g-i = 0.68$) and has a disturbed morphology raising the possibility that it is infalling into the cluster for the first time (see Sec. \ref{sec:Discussion} for confirmation of this). We visually determined this UDG to be GC-poor. In Figure \ref{fig:gc_rich_poor}, we display composite colour images centred on each Perseus UDG in Figure \ref{fig:gc_rich_poor}. Here, the UDGs are labelled `GC-poor' and `GC-rich' as appropriate. For the remainder of the paper we drop the `PUDG\_' prefix for each Perseus UDG.

\subsection{Detailed Fitting}
To perform more detailed 2-D image fitting, we generated $2 \times 2$ arcmin$^2$ cutout images around each object in both $g$- and $i$-bands. These cutouts were large enough to provide sufficient area to reliably estimate the background level. We also generated cutouts of the variance map from \texttt{hscPipe} \citep{HSC_pipe} and used them as the per-pixel flux uncertainties in the fitting. The point spread function (PSF) model of each object was reconstructed for the centre of the cutout based on \texttt{PSFex} results \citep{Bertin2011}. Using \texttt{SEP} \citep{Bertin1996, Barbary2016}, we detected and masked out all objects on the cutout that were not associated with the UDG's stellar body. This was challenging for R16 as it is surrounded by a large amount of Galactic cirrus and it was not possible to fully mask some of the very low surface brightness cirrus. S74 is close to a saturated star and a bright background galaxy. For this UDG we slightly loosened the masking criteria to ensure the main stellar body of S74 was not masked. We experimented with different masking strategies and determined our choice of final masking does not affect the fitting results.

After masking and PSF convolution, we fitted each UDG as a two-dimensional {S\'{e}rsic} function \citep{Sersic1968} using the multi-component galaxy image fitting tool \texttt{Imfit} (\citealp{Erwin2015}\footnote{https://www.mpe.mpg.de/~erwin/code/imfit/}, similar to e.g., \citealp{Roman2017b, Roman2019, Roman2021, kadofong2020, Saifollahi2021}).  R15, R84, and S74 all show clear point source-like objects at their centres (Fig.~\ref{fig:gc_rich_poor}) that needed to be modelled simultaneously to the rest of the UDG. We included these nuclei as a two-dimensional Gaussian component in our model. To achieve a robust model, we adopted the Nelder--Mead simplex algorithm for $\chi^2$ minimisation to avoid any possible local minima found in the fitting process. At the end of each fit, we used the residual image to verify that model adequately accounts for the flux distribution of a UDG. We display an example fit for R15 in Figure~\ref{fig:r15_imblock}. 

For low surface brightness galaxies, an incorrect sky background estimate can significantly alter the model (see e.g., Appendix C of \citealp{Pandya2018}). We therefore also tested our sky background model by modelling it as a ``tilted plane" with gradients in both X and Y directions. The result of this test indicated that the absence of a background component in our fitting did not affect our final results.

Although the HSC $g$-band data are used primarily in this work, fitting was performed in all three $g$,$r$,$i$-bands. Fitting in each band was found to be consistent with one another. Furthermore, fitted values obtained from the HST imaging of the UDGs R15, R16 and R84 are in general agreement with the values reported herein. These tests provide additional confidence to the robustness of our fitting process.

The $\chi^2$ minimisation algorithm does not usually provide meaningful uncertainties of the model parameters when fitting with \texttt{Imfit}. We therefore used the built-in capability of \texttt{Imfit} to perform 1000 bootstrap fitting iterations on images resampled based on the provided noise level. We used these bootstrap fits to estimate uncertainties for our best-fit results. We summarise the results of our more extensive fitting process, along with associated uncertainties, in Table~\ref{tab:imaging}.

For conversion of distance dependent data, we assume a distance to the Perseus cluster of 75 Mpc ($m-M$ of 34.38). All magnitudes and surface brightnesses are extinction corrected using \citet{Schlafly2011} and are quoted in the AB system. We note the maps of \citet{Schlafly2011} are significantly coarser than our data and thus a systematic error may be introduced when applying these dust corrections to our data. For $g-i$ colours we subtract extinction corrected magnitudes from fitting the HSC $i$-band data from our $g$-band fits. We calculate stellar masses using UDG luminosities and their $g-i$ colour in the $M_{\star}/L_{g}$ relation of \citet{Into2013} (i.e., $M_{\star}/L_{g}$ min 1.1, max 3.1).

\subsection{Sample Characteristics}
\begin{figure}
    \centering
    \includegraphics[width = 0.46 \textwidth]{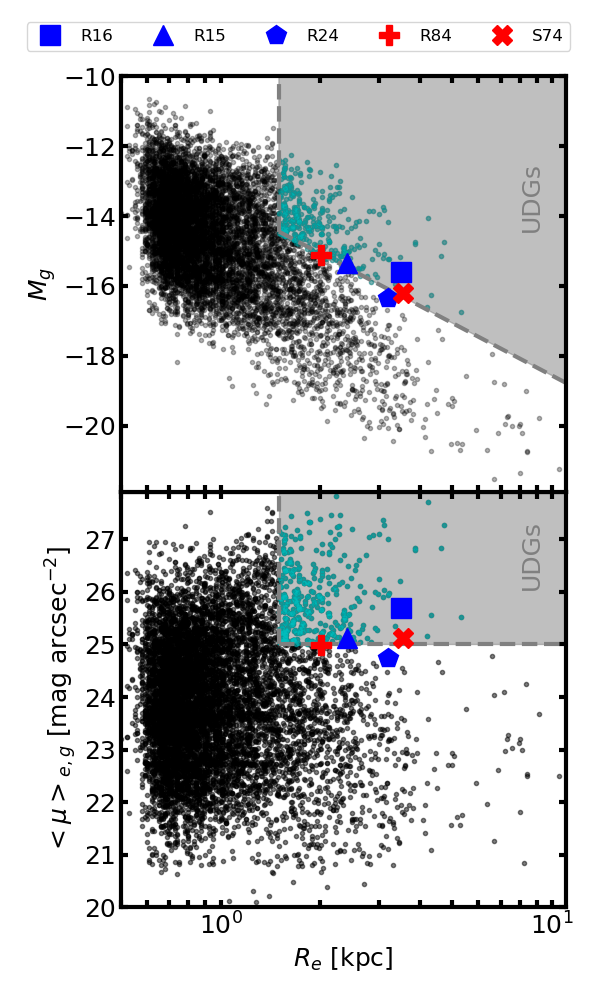}
    \caption{$g$-band magnitude (\textit{top panel}) and average surface brightness within half-light radius (\textit{bottom panel}) \textit{vs.} half-light radius. We plot our Perseus UDGs against the $V$-band catalogue of Coma cluster galaxies from \citet{Alabi2020} (black points). Galaxies in the \citet{Alabi2020} catalogue meeting the UDG definition are plotted in cyan. The parameter space where UDGs reside has been highlighted by the grey region. Our Perseus sample exists at the borderline of UDG parameter space with magnitudes and surface brightnesses making them amongst the brightest UDGs. R24 is slightly too bright to reside in UDG parameter space but is expected to fade into the UDG regime with quenching.}
    \label{fig:context}
\end{figure}

Noting the original UDG definition of \citet{vanDokkum2015}  (i.e., $\mu_{0,g}>24\ \mathrm{mag\ arcsec^{-2}}$ and $R_{\rm e} > 1.5$ kpc) is affected by UDGs with a central nucleus, many literature works instead choose the average surface brightness within the half-light radius ($\langle\mu\rangle_{e,g}$) when defining UDGs (e.g., \citealp{Yagi2016, vanderBurg2017, Greco2018, Janssens2019}). When discussing the UDG criterion in terms of $\langle\mu\rangle_{e,g}$ it is important to also add $\sim$ 1 mag to the definition to allow for the larger aperture being probed in order to ensure the galaxies are not brighter than those of the original definition. We therefore adopt a UDG definition of $\langle\mu\rangle_{e,g}>25\ \mathrm{mag\ arcsec^{-2}}$ and $R_{\rm e} > 1.5$ kpc in this work. 

In Figure \ref{fig:context} we contextualise our Perseus cluster UDG sample in relation to galaxies from the Coma cluster \citep{Alabi2020}. Under the assumption of Perseus cluster membership (see Sec. \ref{sec:Discussion} for a confirmation of this) our UDGs are amongst some of the biggest and brightest currently observed. This is a simple selection effect whereby it was necessary to observe the brightest galaxies in order to ensure the data had sufficient S/N to be usable. We note R24 does not formally meet the UDG definition, having a $\langle\mu\rangle_{e,g}$ $\sim$ 0.25 mag too bright. Given its blue colour and visually disturbed morphology, it is likely beginning its infall into the Perseus cluster and will be expected to quench and fade into the UDG regime within $\sim$ 0.5 Gyr(see e.g., \citealp{Roman2021} fig. 4). 

\section{KCWI Data} \label{sec:KCWI_Data}

\subsection{KCWI Acquisition}\label{sec:KCWI_aquisition}
\begin{figure*}
    \centering
    \includegraphics[width = 0.97 \textwidth]{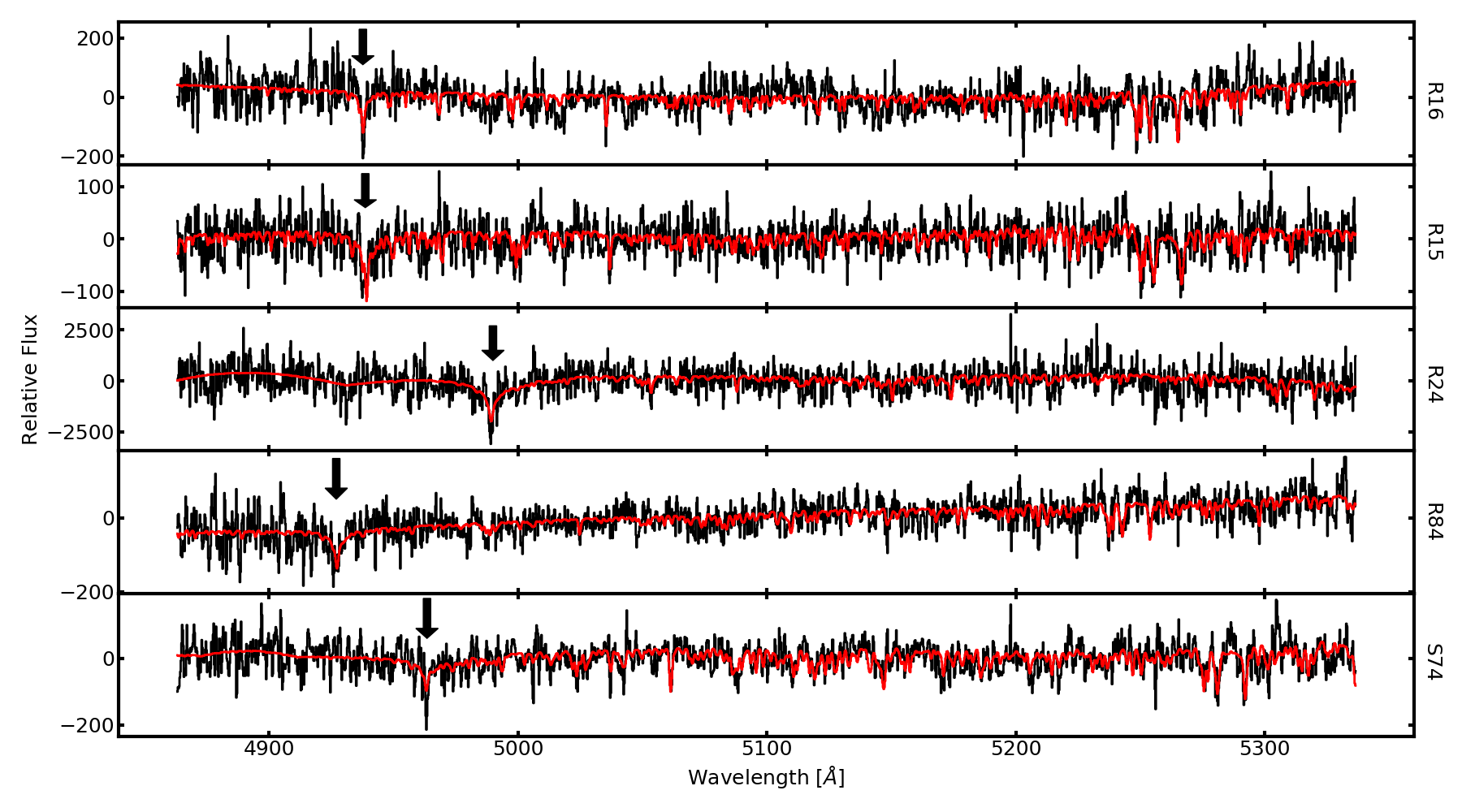}
    \caption{The five KCWI spectra analysed in this work. We plot the final spectra for each UDG (black) along with a representative \texttt{pPXF} fit from our extensive fitting process (red). Perseus UDG names for each spectra are labelled on the right of each panel. In each spectrum we indicate the positioning of H$\beta$ with a black arrow. Based on fits such as these we measure the recessional velocities and velocity dispersions as reported in Table \ref{tab:udg_spec}.}
    \label{fig:spectra}
\end{figure*}

\begin{table}
    \centering
    \begin{tabular}{ccccc}
    \\ \hline
    Name & S/N & $V_{r}$ & $\sigma$ & $M(<R_{1/2})$ \\
     PUDG\_ & [\AA$^{-1}$] & [$\mathrm{km\ s^{-1}}$] & [$\mathrm{km\ s^{-1}}$] & [$\mathrm{\times 10^{8}\ M_{\odot}}$] \\ \hline
    R16 & 12 & 4679 (2) & 12 (3) & 4.70 (2.41) \\
    R15 & 15 & 4762 (2) & 10 (4) & 2.25 (1.82) \\
    R24 & 14 & 7784 (4) & 20 (5) & 11.9 (6.19) \\
    R84 & 18 & 4039 (2) & 19 (3) & 6.74 (2.16) \\
    S74 & 16 & 6215 (2) & 22 (2) & 16.0 (3.03) \\
    \hline
    \end{tabular}
    \caption{A summary of the properties of our Perseus UDG sample. Columns from left to right are as follows: 1) Designation used throughout this work; 2) Signal-to-noise ratio on the continuum; 3) UDG recessional velocity; 4) UDG velocity dispersion and 5) Dynamical mass within 3D, circularised half-light radius. Kinematic properties are measured from the KCWI spectra described in Section \ref{sec:KCWI_Data}. When relevant, uncertainties are given in (brackets) after values.}
    \label{tab:udg_spec}
\end{table}

\begin{figure*}
    \centering
    \includegraphics[width = 0.85 \textwidth]{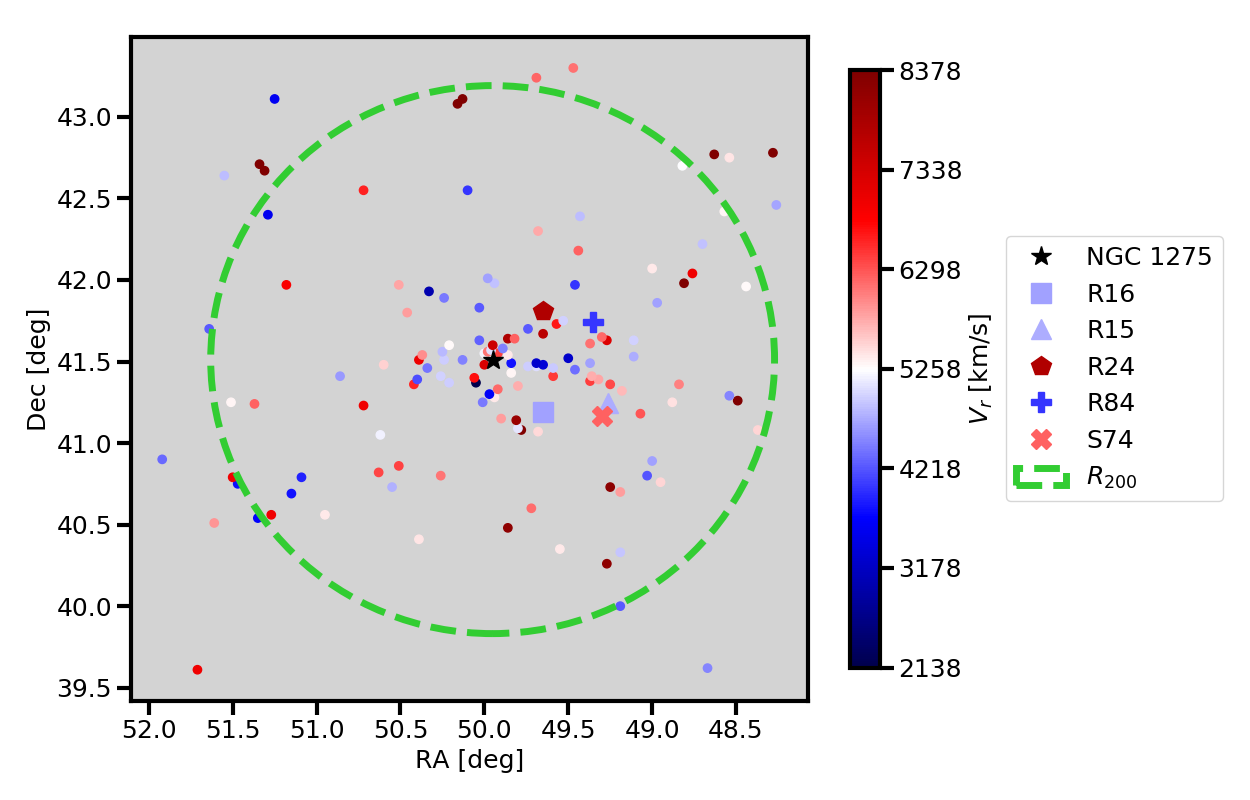}
    \caption{UDG positioning within the Perseus cluster. We plot our UDGs (R15 - triangle, R16 - square, R25 - pentagon, S74 - cross and R84 - plus) along with other, brighter galaxies in the direction of the Perseus cluster from the 2M++ sample of \citet{Lavaux2011} (circles). We include the central, bright cluster galaxy NGC 1275 as a reference (black star). We colour code points by their difference from the mean recessional velocity of the Perseus cluster. Our colour scheme is set to have upper and lower limits of 3 times the Perseus galaxy velocity dispersion about the mean recessional velocity of the cluster. A green circle is included to represent the virial radius of the cluster. All UDGs are projected to be well within the cluster limits.     }
    \label{fig:cluster_pos}
\end{figure*}

The integral field spectroscopy detailed in this work was acquired using the Keck Cosmic Web Imager (KCWI; 
\mbox{\citealp{Morrissey2018}}) during observing runs on the nights 2018 November 12/13, 2018 December 12/13, 2019 October 29/31, 2020 October 20/21, 2020 November 11 and 2020 December 13. We used the medium slicer and the BH3 grating (R $\approx$ 9750). Observations were largely conducted in dark, clear conditions with good seeing ($<1.5$''). An exception were the nights of 2020 November 11 and 2020 December 12, where the seeing was poor ($>2.5$''). We note that we `light-bucket' collapsed our observations of the UDGs so poor seeing is not expected to affect our science results. We also note the nights 2019 October 31, 2019 November 1, 2020 October 20 and 2020 October 21 which had varying levels of high altitude cloud throughout. We found a large decrease in signal-to-noise in our observations of R24 due to this cloud. In order to remove affected frames we visually inspected the whitelight depictions of our data cubes to ensure that R24 was visible, removing those where cloud cover was evident. We then took the remaining frames and coadded in frames in two stages, ensuring each stage increased the signal-to-noise to the final stack and removing those that did not. Of the 20 1200s exposures observed targeting R24, 8 were used in our final stack. We summarise our observations, including observing conditions, program ID and integration times, in Table \ref{tab:obs_summary} of Appendix \ref{app:obs_summary}. Final exposure times were: 25200s for R15, 48600s for R16, 9600s for R24, 25533s for R84 and 20400s for S74. 

The data were initially reduced using the KCWI data reduction pipeline (KDERP). Following this we took the non-sky subtracted, differential atmospheric refraction corrected and standard star calibrated `ocubes' and performed the trimming and extra flat fielding steps described in \citet{Gannon2020}. 

R15 and R84 are sufficiently small that we were able to target the galaxies with the medium slicer and include on-chip sky. Observations of R15 were taken in split configurations, dithering between having sky observed on the CCD above and below the UDG. For sky observed below the UDG, we extract spectra using a 21 $\times$ 40 spaxel box centred on the target, taking the rest of the slicer as sky. For sky observed above the UDG, we extracted a 21 $\times$ 33 spaxel box centred on the UDG, taking the rest of the slicer as sky. In the case of R84 we extracted a 21 $\times$ 24 spaxel box centred on the UDG taking the rest of the slicer as sky.

R16, R24 and S74 are of sufficient size to fill the field of view of the medium slicer and so offset sky observations are used to perform sky subtraction using the technique described in \citet{Gannon2020}. Briefly, we use a template for galaxy emission along with an ensemble of offset sky observations to model the non-sky subtracted spectrum and perform sky subtraction. For each of the galaxies, we take the 10 sky frames observed temporally nearest to the science observation being subtracted and use them in this modelling. We note some of these sky exposures may come from other nights or targets during our Perseus observations.  

Once the data were sky-subtracted, relevant barycentric corrections were applied \citep{Tollerud2013} and they were median stacked. The resulting spectra have signal-to-noise per \AA~of: 15  for R15, 12 for R16, 14 for R24, 16 for S74 and 18 for R84. These signal-to-noise levels are all quoted on the continuum (Table \ref{tab:udg_spec}). 

\subsection{Spectral Fitting}
In order to extract stellar kinematics from our spectra we fitted the spectra using the method described in \citet{Gannon2020} which utilises the code \texttt{pPXF} \citep{Cappellari2017}. Briefly, the spectrum was fitted with the high resolution \citet{Coelho2014} synthetic spectral library using 241 differing input configurations for \texttt{pPXF}. Fits that reported uncertainties greater than 25 $\mathrm{km\ s^{-1}}$ in either recessional velocity or velocity dispersion were discarded. Our final reported values were taken from the median of the remaining fits with uncertainties taken from the $1$-sigma spread in these values. We provide an example of a fit we deem acceptable for each galaxy in Figure \ref{fig:spectra}. We consistency-checked our fitting process by splitting the spectra into red and blue halves and fitted these also. Additionally, we performed the same fitting using a KCWI observation of the Milky Way globular cluster M3 as a template.

R84 is the only UDG where the additional regions and template fits did not report recessional velocities within 1 pixel ($\sim$ 13 $\mathrm{km\ s^{-1}}$). Here the blue-half of the spectrum reported recessional velocities and velocity dispersions in strong disagreement with the other fitting. We note that this spectrum has noticeably weaker iron absorption features than the other 4 UDGs (Figure \ref{fig:spectra}) and so fitting in the blue half is dominated by the H$\mathrm{\beta}$ line, which poorly constrains the fit on its own. In the fitting of R15 with the M3 template, all three spectral regions (red-half/blue-half/all) yielded higher velocity dispersions than when fitted with the \citet{Coelho2014} library. This was most likely caused by numerical biases that can become present when trying to recover velocity dispersions around the resolution limits of the template(s) being used \citep{Robertson2017}. The velocity dispersion reported by the \citet{Coelho2014} library is below this resolution limit for the M3 template. Final values from the spectral fitting for each of the 5 galaxies are reported in Table \ref{tab:udg_spec}. 

An additional sky subtraction on S74 was performed using the on-chip `sky'. Here we subtracted the fainter outskirts of the galaxy from the central, brighter regions to produce a spectrum and confirm the results of our offset sky subtraction using the \citet{Gannon2020} sky subtraction routine. We stacked the sky subtracted spectra and then fit in the same manner as for the main sample. The final recessional velocity ($6217\pm 2$ $\mathrm{km\ s^{-1}}$) and velocity dispersion ($21\pm3$ $\mathrm{km\ s^{-1}}$) agree well with the values reported for our on chip sky subtraction of S74.

\subsection{Dynamical Mass}
Strictly speaking, to infer dynamical masses we need to measure the luminosity-weighted line-of-sight velocity dispersion within the half-light radius. As our galaxies have half-light radii of $\sim$ 6'' to 11'', their effective diameters are well matched to the size of KCWI's medium slicer ($\sim$16'' $\times$ 20''). Our velocity dispersions will therefore be approximately luminosity-weighted within the half-light radius.

We measure dynamical masses within the 3D de-projected half-light radius ($R_{1/2}$) using the mass estimator of \citet{Wolf2010}. Using the 2D projected, circularised half-light radius ($R_{e, \mathrm{circ}}$) and the luminosity-weighted line-of-sight velocity dispersion within this radius ($\sigma$), it takes the form: 

\begin{equation} \label{eqtn:wolf}
M(<R_{1/2}) = 930 \left( \frac{\sigma_{e}^{2}}{\mathrm{\left(km\ s^{-1}\right)^{2}}}\right) \left(\frac{R_{e, \mathrm{circ}}}{\mathrm{pc}}\right)\ \mathrm{M_{\odot};}\quad
\end{equation}
\begin{equation*}
\mathrm{where}\ R_{1/2} \approx \frac{4}{3} R_{e, \mathrm{circ}}
\end{equation*}

We list calculated dynamical masses for each UDG, along with associated uncertainties, in Table \ref{tab:udg_spec}.

We caution that there exists evidence that the \citet{Wolf2010} formula may become biased in the case of dispersion supported systems undergoing tidal disruption \citep{Errani2018, Carleton2018}. R24 displays a non-smooth stellar body (see Fig. \ref{fig:gc_rich_poor}) which suggests it may be undergoing a tidal interaction and thus have a biased mass estimate from Equation \ref{eqtn:wolf}. In principle, this might affect our other Perseus UDGs as tidal features (if present) are likely below the surface brightness limit of our data due to their fast diffusion times \citep{Penarrubia2008, Carleton2018}. We note however, the work of \citet{Doppel2021} who suggest biases due to tidal effects are short lived in dwarf galaxies after pericentric passages. 

\section{Discussion} \label{sec:Discussion}

\begin{table}
\begin{tabular}{llllll}
\\ \hline
Infall Region & & \multicolumn{4}{l}{Infall Time Period {[}Gyr{]}} \\
 & PI & 0\textless{} & 3.63\textless{} & 6.45\textless{} & Interloper \\
 & & \textless{}3.63 & \textless{}6.45 & \textless{}13.7 & \\
 \hline
Yet to & 0.35 & 0.04 & 0.05 & 0.01 & 0.54 \\
Late & 0.02 & 0.37 & 0.15 & 0.31 & 0.14 \\
Mixed Times & 0.13 & 0.32 & 0.20 & 0.23 & 0.12 \\
Early  & 0.18 & 0.20 & 0.27 & 0.17 & 0.19 \\
Very Early & 0.03 & 0.17 & 0.22 & 0.52 & 0.06\\
\hline
\end{tabular}
\caption{Fractions of galaxies in each infall time period for the coloured regions of Figure \ref{fig:r_rvdiff} from the simulations of \citet{Rhee2017} (rounded to 2 decimals). Regions are named corresponding to their dominant population of galaxies. `PI' is an acronym for `pre-infall'. Interlopers are galaxies that are non-members of the cluster despite projection effects placing them within projected cluster phase space.}
\label{tab:infall_summary}
\end{table}

\begin{figure*}
    \centering
    \includegraphics[width = 0.95 \textwidth]{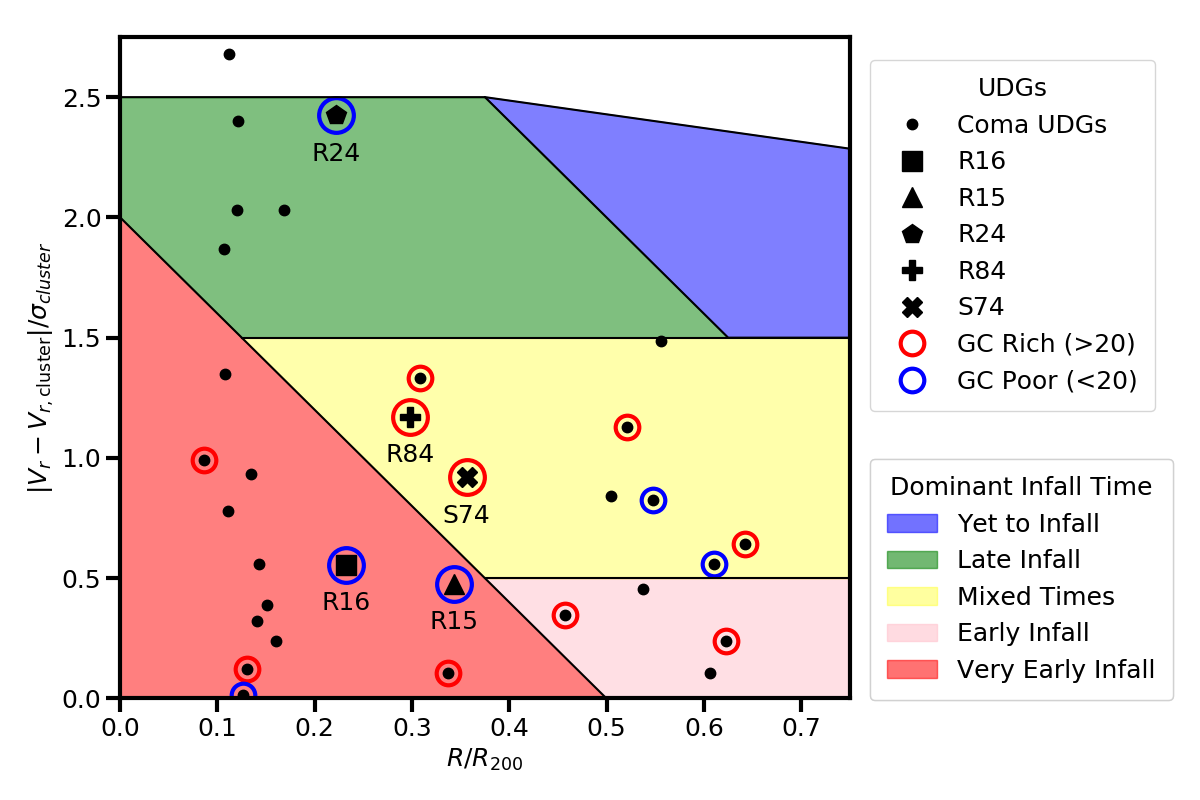}
        \caption{A phase space diagram of the Perseus and Coma cluster UDGs. We plot the absolute difference of the galaxy recessional velocity from the mean of the cluster, normalised by the cluster velocity dispersion \textit{vs.} projected radius from cluster centre (normalised by the virial radius of the cluster). We plot our Perseus UDGs (triangle, square, pentagon, cross and plus) along with UDGs in the Coma cluster (\citealp{Alabi2018} and Yagi358 from Gannon et al. in prep.). UDGs hosting GC systems measured in excess of 20 GCs are deemed GC-rich (circled in red). UDGs hosting GC systems with fewer than 20 GCs are deemed GC-poor (circled in blue). Coma UDG GC system estimates are from \citet{Forbes2020} and references therein. Perseus GC richness categories are as designated by this work. We overlay regions from the simulations of \citet{Rhee2017} to aid in our interpretation of the UDGs (red/pink/yellow/green/blue regions). We note our phase space diagram is in 2D, projected space and that the 3D radius and/or velocity for objects may be much larger. Our Perseus UDG sample appears well bound to the cluster. It is more likely that our Perseus GC-poor UDGs have undergone cluster infall at earlier times than their GC-rich counterparts. This effect is diluted when adding in the Coma cluster to our analysis.}
    \label{fig:r_rvdiff}
\end{figure*}

\begin{figure}
    \centering
    \includegraphics[width = 0.49 \textwidth]{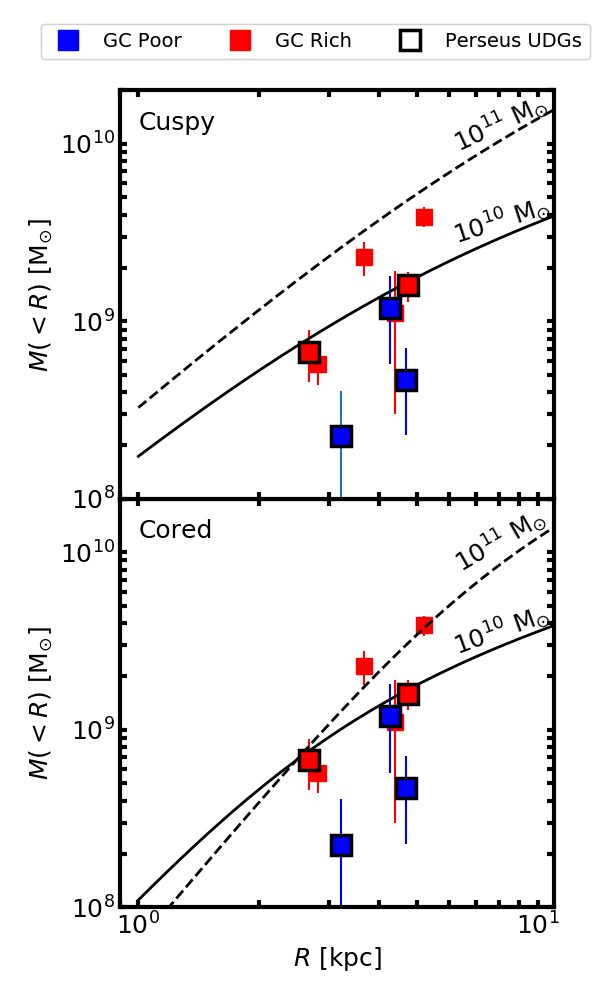}
    \caption{Enclosed mass \textit{vs.} galactocentric radius. We plot the dynamical mass measurements from stellar kinematics for our Perseus UDGs (black outline) along with those from the literature \citep{vanDokkum2017, vanDokkum2019b, Gannon2020, Forbes2021}. UDGs are colour coded by GC richness with red points GC-rich and blue GC-poor. \textit{Top panel:} we compare to cuspy, NFW \citep{Navarro1996} halos of mass $1 \times 10^{10} \mathrm{M_{\odot}}$ (solid line) and $1 \times 10^{11} \mathrm{M_{\odot}}$ (dashed line). \textit{Bottom panel:} we compare to cored \citep{DiCintio2014} halos of mass $1 \times 10^{10} \mathrm{M_{\odot}}$ (solid line) and $1 \times 10^{11} \mathrm{M_{\odot}}$ (dashed line). Overall, GC-rich UDGs have higher dynamical masses than those that are GC-poor. When trying to connect this to their total halo mass the interpretation is more complex (see text for details).
    }
    \label{fig:mdyn_mhalo_compare}
\end{figure}

\begin{figure*}
    \centering
    \includegraphics[width = 0.98 \textwidth]{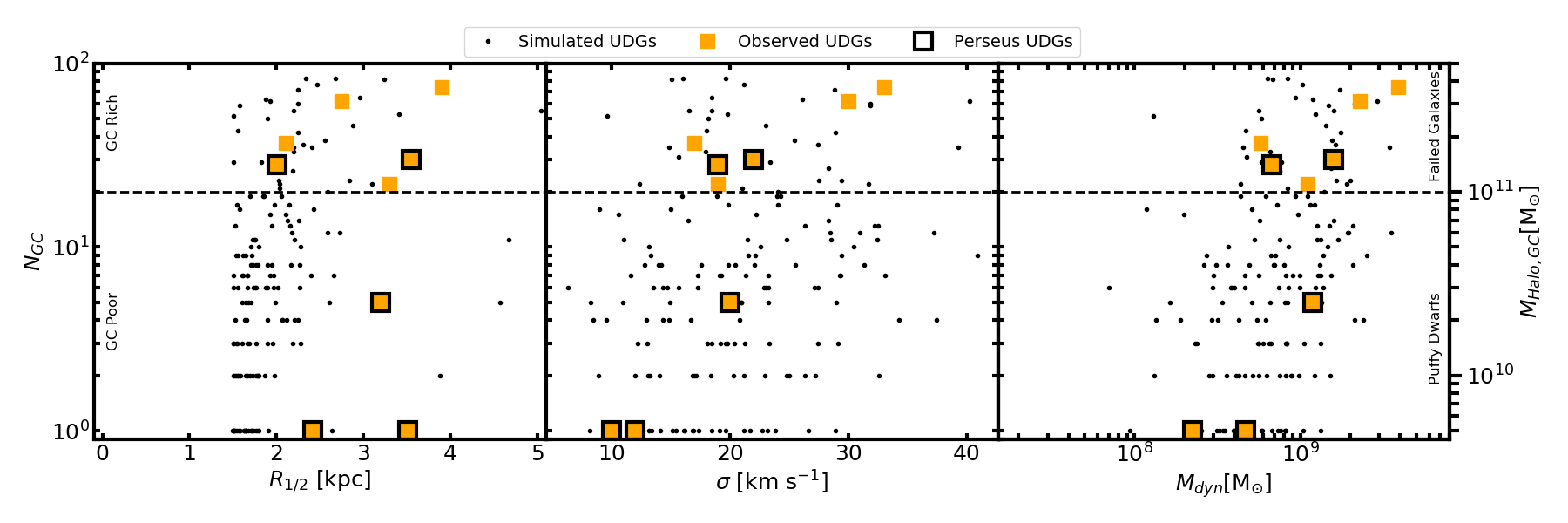}
    \caption{GC number, a known halo mass indicator for normal galaxies, \textit{vs.} half-light radius (\textit{left panel}), stellar velocity dispersion (\textit{centre panel}) and dynamical mass (\textit{right panel}). We include a second y-axis of total dark matter halo masses using the $N_{\rm GC}$ -- $M_{\mathrm{Halo}}$ of \citet{Burkert2020}. We plot the three Perseus UDGs with approximate GC numbers (R15, R16 and R84) from Janssens et al. (in prep.) along with a sample of UDGs from the literature (orange squares; \citealp{Beasley2016,  Cohen2018,  vanDokkum2017, vanDokkum2018, vanDokkum2019, vanDokkum2019b, Gannon2020, Muller2021, Forbes2021}). We add to this sample our two Perseus UDGs without a $N_{\rm GC}$ measurement by assuming a roughly expected GC number. We also plot UDGs from the simulations of \citet[black points]{Carleton2021}. Labels are included for regions of GC richness as prescribed by this work and separated by horizontal dashed lines. ``GC-poor" systems indicate low total halo masses and likely correspond to ``Puffy Dwarf" UDGs. ``GC-rich" systems indicate high total halo masses and likely correspond to ``Failed Galaxy" UDGs. The plot is labelled appropriately for both. Clear observational trends are apparent for UDG GC system richness with both dynamical mass and stellar velocity dispersion.}
    \label{fig:vs_carleton}
\end{figure*}

\begin{figure*}
    \centering
    \includegraphics[width = 0.99 \textwidth]{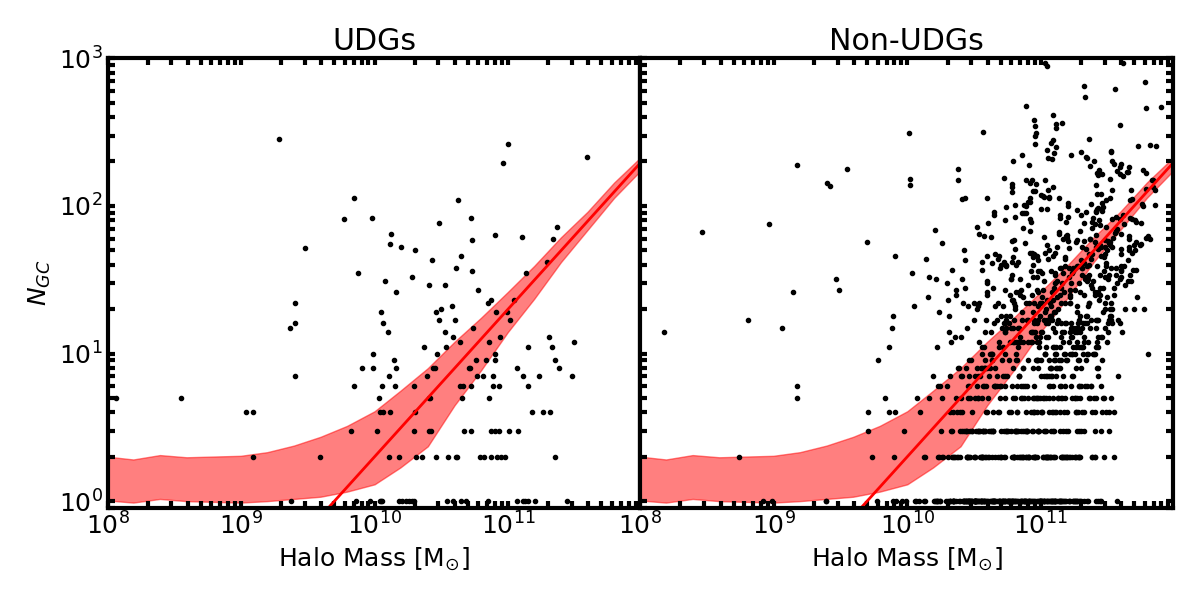}
    \caption{Globular cluster number \textit{vs.} halo mass. We plot \citet{Carleton2021} simulated UDGs (black points; \textit{left}) and the simulated data for a stellar mass matched sample of non-UDG galaxies in the UDG stellar mass range (black points; \textit{right}). We include the observationally established relation of \citet{Burkert2020} (red line) for non-UDGs with indicative scatter (red fill). The simulations of \citet{Carleton2021} do not reproduce the observed $N_{\rm GC}$ -- $M_{\mathrm{Halo}}$ relationship even for non-UDGs.}
    \label{fig:concerns}
\end{figure*}

\subsection{Confirmation of UDG Status}

In Figure \ref{fig:cluster_pos} we plot the positions on the sky of our Perseus UDGs, along with galaxies from the 2M++ sample of relatively bright galaxies \citep{Lavaux2011} that are in the direction of the Perseus cluster. All of our UDGs have recessional velocities within 2.5$\sigma$ of the mean recessional velocity of the Perseus cluster. Additionally, all of our UDGs are well within 0.2-0.4 $R_{200}$ (projected) of the Perseus cluster. It is therefore highly likely that all of our Perseus UDGs are cluster members (see Figure~\ref{fig:r_rvdiff} for an alternate visualisation of this). This environmental association is fully consistent with our assumption of a 75 Mpc distance for our Perseus UDGs. This distance formalises their large half-light radii ($R_{\rm e}>1.5\ \mathrm{kpc}$). Combining this with the surface brightness measurements in Section~\ref{sec:HSC_Data} and Table~\ref{tab:imaging} we confirm the status of our sample as four `bona fide' Perseus UDGs and one galaxy, R24, which is likely soon to fade into the UDG regime.

\subsection{UDG Infall Times}\label{sec:infall}

In Figure \ref{fig:r_rvdiff} we plot the difference in galaxy recessional velocity with the mean of the cluster normalised by the cluster's velocity dispersion \textit{vs.} the projected radius from the centre of the cluster in units of cluster virial radii. We supplement our Perseus UDG sample with UDGs from the Coma Cluster (\citealp{Alabi2018} and the UDG Yagi358, $V_{r} = 7967\ \mathrm{km\ s^{-1}}$, from Gannon et al. in prep.; \citealp{Yagi2016}). For the Coma cluster UDGs we colour code them according to their measured GC system richness from \citet{Forbes2020} and references therein. For our Perseus UDGs we similarly colour code them on the basis of their identified GC richness class.

In order to quantify the likely infall timescales of the UDGs we over-plot the phase-space regions derived from galaxies located in 16 clusters within the cosmological, hydrodynamic simulations of \citet{Rhee2017}. These regions represent the different time periods in which galaxies have most likely fallen into the cluster and are designed for easy comparison to observations. We note figure 6 of \citet{Aguerri2020} provides empirical observational evidence that other Perseus galaxies are broadly consistent with the \citet{Rhee2017} predictions. We summarise the expected fractions of galaxy infall times in each region based on the simulations of \citet{Rhee2017} in Table \ref{tab:infall_summary}. 

For example, the red region plotted in Figure \ref{fig:r_rvdiff} is dominated by galaxies that have fallen into the cluster at the earliest times. More specifically $\sim 52\%$ of galaxies found in this region are expected to have infall times $>6.45$ Gyr ago. Of the remaining galaxies, we expect $\sim 22\%$ to have infall times between 3.63 and 6.45 Gyr and $\sim 17 \%$ to have only recently fallen into the cluster ($<3.63$ Gyr ago). The remaining, `interloper', galaxies are not located in the cluster despite projection effects placing them within cluster phase space.


The UDGs in our Perseus sample all have phase-space locations consistent with them being bound to the Perseus cluster. An exception may be R24 which is located in the green region of Figure \ref{fig:r_rvdiff}. This suggests that if it is not infalling for the first time it is probable that it will fall in soon. Of interest for the other 4 UDGs is the difference between the two GC-rich and two GC-poor UDGs.

Our data suggest that the GC-poor UDGs (R15 and R16) are more likely to have fallen into the cluster at earlier times than the GC-rich UDGs (S74 and R84). However, we caution that we are dealing with small number statistics and that, even within the regions from the simulations of \citet{Rhee2017}, there are significant variations in the infall times of individual galaxies. The observed trend largely disappears after the inclusion of data from the Coma cluster.

Formation scenarios where GC-rich UDGs form preferentially at high-redshift and quench early (e.g., \citealp{Carleton2021}) logically require the early accretion of GC-rich UDGs to the cluster for the quenching to be environmentally driven. Objects accreted to the cluster at early times are expected to have smaller relative velocities and orbital radii which reflect the smaller gravitational potential at the time of accretion (see further \citealp{Rhee2017}). Therefore, if GC-rich UDGs are environmentally quenched at high redshift we have an expectation that they will inhabit the more bound regions (red/pink) of Figure \ref{fig:r_rvdiff}.

Conversely, GC-poor UDGs are expected to have formation pathways and dark matter halo masses similar to field low surface brightness dwarfs (e.g., \citealp{Beasley2016b}). Due to this they cannot survive long in the strong tidal fields of the cluster core and are expected to predominantly inhabit the later earlier infall times regions of Figure \ref{fig:r_rvdiff} (blue, green, yellow). We therefore expect a trend between UDG infall time and UDG GC system richness where GC-Rich UDGs infall at earlier times. 

The lack of this trend in Figure \ref{fig:r_rvdiff} suggests that GC-rich UDG quenching is not environmentally driven.  Furthermore, the axial alignment of UDGs at radii $> 0.5 R_{200}$ of the cluster Abell~2634 suggests they have experienced little evolution in the cluster thus far \citep{Rong2020}. If UDG quenching is not environmentally driven it may be either self induced or triggered via pre-processing of GC-rich UDGs in a group environment. We note, it has already been observed for some UDGs that quenching episodes can precede their cluster infall (e.g., M-161-1 \citealp{Papastergis2017}; Dragonfly 44 \citealp{Alabi2018}) implying environmental processes are not the only driver of their evolution. Further constraints on these mechanisms can be added through the analysis of their star formation histories (see further \citealp{Ferre-Mateu2018}). For our Perseus UDGs this will be presented in another paper (Ferr\'e-Mateu, in prep.). 

\subsection{Perseus UDG Halo Masses} \label{sec:pudg_halo_masses}

In Figure \ref{fig:mdyn_mhalo_compare} we compare the dynamical mass measurements for our Perseus UDGs to cuspy, NFW dark matter halos \citep{Navarro1996} and cored dark matter halos  \citep{DiCintio2014}. We include four literature UDGs with stellar kinematics and GC counts \citep{vanDokkum2017, vanDokkum2019b, Gannon2020, Forbes2021}. All are GC rich. When considering the four Perseus galaxies that currently meet the UDG criterion (i.e., not R24) along with the literature UDGs, we find that UDGs with a rich GC system have greater dynamical mass within their 3D half-light radius. This agrees with a first order interpretation of GC-rich UDGs residing in higher mass dark matter halos.

Due to our small sample of UDGs we ran simulations to see how often we expect this trend to arise via random chance. For each simulation we randomly assigned either a high dynamical mass or low dynamical mass to the 9 UDGs (5 Perseus + 4 literature) with equal probability. We ran 10 000 simulations. In only 78 simulations of the 10 000 (0.78\%) did we observe the trend seen in Figure \ref{fig:mdyn_mhalo_compare}. Namely, in these simulations we observe all GC-rich UDGs to have high dynamical masses while 2 of 3 GC-poor UDGs had low dynamical masses. If we exclude the GC-poor UDG with high dynamical mass, R24, which both does not currently fit the UDG criteria and may not be in dynamic equilibrium, only 47 of our 10 000 simulations (0.47\%) match the trends seen in Figure \ref{fig:mdyn_mhalo_compare}. We conclude it is highly unlikely that our observation of GC-rich UDGs having higher dynamical masses than GC-poor UDGs has arisen based on random fluctuations alone.

When taking into account the effect of extrapolating dynamical masses into total halo masses using halo profiles, the interpretation of Figure \ref{fig:mdyn_mhalo_compare} becomes more complex. Greater dynamical mass implying greater total halo mass is only generally true under the assumption of cuspy dark matter halos for UDGs with fixed halo parameters (e.g., concentration parameters drawn from the same relation such as \citealp{Dutton2014}). Unfortunately, it is likely misleading to compare UDG dynamical masses to cuspy dark matter halo profiles. There is evidence that favours cored halo profiles for UDGs in simulations \citep{DiCintio2017, Carleton2018, Martin2019} and observations \citep{vanDokkum2019b, Wasserman2019, Gannon2020}. 

In the case of cored halos, lone dynamical masses at a single radius poorly constrain the total halo mass of the UDG (see further \citealp{Gannon2021}). For example, it is clear that our dynamical mass measurement for R84 agrees well with cored halos of both mass $10^{10}\ \mathrm{M_\odot}$ and $10^{11}\ \mathrm{M_\odot}$ and so cannot constrain the total halo mass of the UDG (Figure \ref{fig:mdyn_mhalo_compare}, \textit{bottom panel}). Thus, when assuming a cored dark matter halo profile our measurement of greater dark matter mass within the central region implies little more than that GC-rich UDGs have greater dark matter mass in their centres. For example, it is possible that they reside in dark matter halos of the same total mass as GC-poor UDGs, but the GC-poor UDGs are more efficient at creating a core and therefore have less mass in the halo centre.

We are therefore able to draw two possible conclusions from Figure \ref{fig:mdyn_mhalo_compare}: 1) GC-rich UDGs reside in halos of greater total mass than GC-poor UDGs, as expected given the GC number -- halo mass relation of \citet{Burkert2020}, or 2) GC-rich UDGs reside in halos of a similar mass to GC-poor UDGs but with a different halo profile. In order to have the same total mass the halo profile must be more concentrated and with higher central dark matter densities. This implies the total dark matter halo is of smaller radius due to the increase in its density. We note option 2) would align with halo concentration predictions for GC-rich UDGs from \citet{Trujillogomez2021} although it is unclear how they are able to reproduce the observed breadth in UDG dynamical masses.

\subsection{Trends with GC System Richness}

In Figure \ref{fig:vs_carleton} we plot our observations for R15, R16 and R84, supplemented by UDGs from the literature \citep{Beasley2016,  Cohen2018,  vanDokkum2017, vanDokkum2019b, Gannon2020, Muller2021, Forbes2021}. R15 and R16 are plotted as having 1 GC as their selected GC counts are consistent with background levels. We also include the Perseus UDGs R24 and S74 that do not have exact $N_{\rm GC}$ measurements by assuming a GC count based on their visually identified class of GC richness as per Section \ref{sec:vis_identify}. For R24 we assume 5 GCs to place it in a representative position within the GC-poor class. For S74 we assume 30 GCs to place it at a similar level of GC richness to R84. 

Literature UDG properties, barring GC counts, are from appendix A of \citet{Gannon2021} and the references therein. For the Virgo cluster UDG VCC~1287 we use the GC system counts of Beasley et al. (2016; $N_{\rm GC} = 22 \pm 8$). For the Coma Cluster UDGs Dragonfly~44 ($N_{\rm GC} = 74 \pm 18$\footnote{Although see \citet{Saifollahi2021} for an alternative view of the GC system richness of Dragonfly 44.}) and DFX1 ($N_{\rm GC} = 62 \pm 17$) we use the measurements of \citet{vanDokkum2017}. For the NGC~5846 group UDG NGC~5846\_UDG1 ($N_{\rm GC} = 37 \pm 5$) we use the GC system measurement of \citet{Muller2021}\footnote{\citet{Muller2021} refers to this UDG as MATLAS-2019.}.

The literature sample of UDGs in \citet{Gannon2021} also includes the UDGs NGC~1052\_DF2 and NGC~1052\_DF4 which have had their GC systems measured. \citet{Shen2021} find evidence that many of the GCs orbiting the NGC~1052 UDGs are actually ultra-compact dwarfs due to their luminosity. The GC systems of both UDGs comprise $\sim 19$ star clusters but it is not obvious how these compare to the other UDGs plotted due to their markedly different luminosity function \citep{Shen2021}. We therefore choose to exclude them when adding the sample of \citet{Gannon2021} literature UDGs to Figure \ref{fig:vs_carleton}. 

We now compare these observed galaxies to the simulated UDGs of \citet{Carleton2021}. They used a semi-analytic model combined with the Illustris-TNG simulation to form GC-rich UDGs in clusters of $M_{200} \ge 2 \times10^{14} \mathrm{M_{\odot}}$. The authors do not quote maximum cluster mass is in the simulation. \citet{Carleton2021} suggests the primary cause of UDG formation in clusters is rapid star formation at high redshift combined with prolonged tidal heating within the cluster's gravitational well. For more details on the UDG formation simulation and the model applied for GC formation see \citet{Carleton2018, Carleton2021}.  

Under the \citet{Carleton2021} simulated UDG formation model there is the expectation that GC number should correlate with half-light radius. However, in the observed UDGs there is no clear trend in GC number -- half-light radius parameter space (Figure \ref{fig:vs_carleton}, \textit{left panel}; Pearson's $r$ = 0.16). The three GC-poor UDGs plotted are from our Perseus sample which is selected to have approximately equivalent half-light radii between GC-rich and GC-poor UDGs. Our observational data is therefore insufficient to comment on the actual trend in half-light radius with GC number. We note that due to the relatively small spread in luminosities for those UDGs with GC number measurements, \citet{Forbes2020} figure 7 provides evidence of a weak inverse trend in this relationship for Coma cluster UDGs. This would be the opposite of the expectation from \citet{Carleton2021}'s simulations. 

Based on the observed $N_{\rm GC}$ -- $M_{\mathrm{Halo}}$ relation of \citet{Burkert2020} there is a first order expectation that GC number should correlate with dynamical mass. For pressure-supported systems it is well established dynamical mass is proportional to the half-light radius and the square of the velocity dispersion (e.g., \citealp{Wolf2010, Errani2018}). We have established that our data are not sufficient to reveal trends with half-light radius. Interestingly, there does seem to be a correlation between GC number and velocity dispersion (Figure \ref{fig:vs_carleton}, \textit{middle panel}; Pearson's $r$ = 0.91). This suggests UDGs with more GCs are dynamically ``hotter". The same trend is not apparent in \citet{Carleton2021} simulated UDGs with many GC-poor UDGs having similar velocity dispersions to GC-rich UDGs.

To first order we expect GC number to correlate with dynamical mass given they correlate with halo mass. Namely, we expect ``failed galaxy'' UDGs to have higher dynamical masses due to their more massive dark matter halo. Similarly, we expect ``puffy dwarf'' UDGs to have smaller dynamical masses indicative of their dwarf-like dark matter halo. We observe this observational trend, between GC number and dynamical mass (Figure \ref{fig:vs_carleton}, \textit{right panel}; Pearson's $r$ = 0.85), although it may be slightly diluted by our choice to half-light radius match in the Perseus sample. A similar trend also appears in the \citet{Carleton2021} simulated UDGs. This is slightly puzzling due to the lack of trend between velocity dispersion and GC number in their simulation.

In Figure \ref{fig:concerns} we investigate the \citet{Carleton2021} simulations further. Here, we compare their simulated galaxies' halo masses and GC numbers to the observationally established $N_{\rm GC}$ -- $M_{\mathrm{Halo}}$ of \citet{Burkert2020}. It is clear neither the simulated UDGs, or the simulated non-UDGs, follow the observational relationship. This is particularly true when considering the tight scatter of the $N_{\rm GC}$ -- $M_{\mathrm{Halo}}$ once a GC system becomes large enough to mitigate statistical effects ($\gtrsim 20$). Particularly troubling are those galaxies hosting large GC systems similar to the Milky Way (i.e., $\gtrsim 150$) but halo masses of order $10^{9}\ \mathrm{M_{\odot}}$. Likewise, there appear to be galaxies of large halo mass (i.e., $\gtrsim 10^{11}\ \mathrm{M_{\odot}}$) but with only 1 or 2 GCs.

Current theory suggests the $N_{\rm GC}$ -- $M_{\mathrm{Halo}}$ relation should be established at $z \gtrsim 6$  \citep{Boylan-kolchin2017}. Our investigations showed the \citet{Carleton2021} simulations do not follow the relationship at cluster infall either, when the $N_{\rm GC}$ -- $M_{\mathrm{Halo}}$ relationship should already be established. We suggest future simulations seeking to simulate the GC systems of UDGs must, as a basic requirement, seek to reproduce the \citet{Burkert2020} relationship for non-UDGs.



\section{Conclusions} \label{sec:Conclusions}
In this work we study five UDGs residing in the Perseus cluster using HSC imaging and KCWI spectroscopy. We find two of them likely harbour GC-rich populations and three have GC-poor populations. Our main conclusions from this study are as follows:

\begin{itemize}
    \item Our KCWI spectroscopy confirms Perseus cluster membership and, along with our HSC imaging, confirms four of our sample of galaxies meet the UDG criteria. The exception is R24 which is slightly brighter than the standard UDG definition. We find this galaxy is likely infalling into the Perseus cluster for the first time and, after quenching, is expected to fade into the UDG regime within $\sim 0.5$ Gyr.
    \item By measuring a stellar velocity dispersion we calculate dynamical masses within half-light radii for our sample. After supplementing our sample with literature UDGs we find UDGs with richer GC systems have greater dynamical mass within the half-light radius. They also have larger velocity dispersions. This is as expected by a first order interpretation that GC-rich UDGs reside in more massive dark matter halos. However, we stress that this does not necessarily indicate GC-rich UDGs \textit{are} in more massive halos. Extrapolation of dynamical masses into total halo masses requires the non-trivial assumption of a dark matter halo profile. A difference in the nature of their halo profiles (e.g., GC-poor UDGs exhibiting a stronger dark matter core) may also explain the difference in central dark matter mass.
    \item No clear correlation exists between GC system richness and phase space positioning in the cluster. This is in contrast to the expectations of high-redshift GC-rich UDG formation scenarios where they are predicted to environmentally quench within clusters at early times. Either internal mechanisms or pre-processing in groups may instead be required to explain the known quenching of UDGs.
    \item No clear correlation exists between UDG GC numbers and host galaxy half-light radii. This is likely due to the lack of an unbiased sample of GC poor UDGs. Our observed GC poor UDGs were selected to have similar sizes to GC rich UDGs.
\end{itemize}

\section{Data Availability}
The KCWI data presented are available via the Keck Observatory Archive (KOA): \url{https://www2.keck.hawaii.edu/koa/public/koa.php} 18 months after observations are taken. The HSC data presented herein will be shared upon reasonable request with the corresponding author. 

\section*{Acknowledgements}

We thank the anonymous referee for their insightful comments which greatly helped improve the quality of the work. We thank A. Alabi and T. Carleton for kindly providing their data for use in Section \ref{sec:Discussion}.JSG and AJR wish to thank A. Alabi for doing preliminary analyses data relating to this work. JSG acknowledges financial support received through a Swinburne University Postgraduate Research Award throughout the creation of this work. AFM has received financial support through the Postdoctoral Junior Leader Fellowship Programme from `La Caixa' Banking Foundation (LCF/BQ/LI18/11630007). AJR was supported by National Science Foundation Grant AST-1616710 and as a Research Corporation for Science Advancement Cottrell Scholar.
Support for Program number HST-GO-15235 was provided through a grant from the STScI under NASA contract NAS5-26555.

The data presented herein were obtained at the W. M. Keck Observatory, which is operated as a scientific partnership among the California Institute of Technology, the University of California and the National Aeronautics and Space Administration. The Observatory was made possible by the generous financial support of the W. M. Keck Foundation. The authors wish to recognise and acknowledge the very significant cultural role and reverence that the summit of Maunakea has always had within the indigenous Hawaiian community.  We are most fortunate to have the opportunity to conduct observations from this mountain.

This paper is based [in part] on data from the Hyper Suprime-Cam Legacy Archive (HSCLA), which is operated by the Subaru Telescope. The original data in HSCLA was collected at the Subaru Telescope and retrieved from the HSC data archive system, which is operated by Subaru Telescope and Astronomy Data Center at National Astronomical Observatory of Japan. This paper makes use of software developed for the Vera C. Rubin Observatory. We thank the observatory for making their code available as free software at  http://dm.lsst.org.




\bibliographystyle{mnras}
\bibliography{bibliography.bib} 



\appendix

\section{KCWI Observations Summary} \label{app:obs_summary}
In the following table we summarise the observing conditions and integration times of the KCWI observations used in this work.

\begin{table*}
\begin{tabular}{cccccc}\label{tab:obs_summary}
UDG & Program ID & Night  & Conditions & Target & Integration Time  \\ \hline

PUDG\_R15 & U216 & 2019 Oct 29 & Clear           & Science & 16 $\times$ 1200s \\
          &      &             &                 & Sky     &  6 $\times$ 1200s \\
          & U216 & 2019 Oct 31 & Sporadic Cloud & Science &  5 $\times$ 1200s \\
          &      &             &                 & Sky     &  2 $\times$ 1200s \\ \hline
          
PUDG\_R16 & U107 & 2018 Nov 12 & Clear           & Science &  6 $\times$ 1200s \\
          &      &             &                 & Sky     &  1 $\times$ 1200s, 1 $\times$ 600s \\
          & U107 & 2018 Nov 13 & Clear           & Science & 13 $\times$ 1200s \\
          &      &             &                 & Sky     &  3 $\times$ 1200s, 1 $\times$ 900s \\
          & U107 & 2018 Dec 12 & Clear           & Science &  9 $\times$ 1200s \\
          &      &             &                 & Sky     &  2 $\times$ 1200s, 1 $\times$ 720s \\
          & U107 & 2018 Dec 13 & Clear           & Science &  3 $\times$ 1800s, 2 $\times$ 1200s \\
          &      &             &                 & Sky     &  1 $\times$ 1200s \\
          & U216 & 2019 Oct 31 & Sporadic Cloud & Science &  6 $\times$ 1200s \\
          &      &             &                 & Sky     &  2 $\times$ 1200s \\ \hline
 
PUDG\_R24 & W140 & 2020 Oct 20 & Strong Cloud   & Science & 11 $\times$ 1200s \\ 
          &      &             &                 & Sky     &  5 $\times$ 1200s, 1 $\times$ 900s \\
          & U088 & 2020 Oct 21 & Mild Cloud     & Science &  9 $\times$ 1200s\\
          &      &             &                 & Sky     &  4 $\times$ 1200s, 1 $\times$ 600s\\\hline

PUDG\_R84 & U088 & 2020 Nov 11 & Poor Seeing, Clear & Science &  8 $\times$ 1200s, 1 $\times$ 813s \\ 
          & U088 & 2020 Dec 13 & Poor Seeing, Clear & Science & 12 $\times$ 1200s, 1 $\times$ 720s\\ \hline

PUDG\_S74 & U216 & 2019 Nov 1  & Sporadic Cloud & Science & 17 $\times$ 1200s \\
          &      &             &                 & Sky     &  5 $\times$ 1200s \\ \hline
\end{tabular}%
\caption{A summary of the KCWI data observed whose analysis comprises part of this work. `Science' observations are those targeting the UDG. `Sky' observations are those targeting offset sky positions.  }
\end{table*}

\bsp	
\label{lastpage}
\end{document}